\documentclass[twocolumn]{aastex631}

\usepackage{times}
\usepackage{amsmath}
\usepackage{natbib}
\usepackage{graphicx}

\begin{document}
\title{High-Frequency Power Spectrum of AGN NGC~4051 Revealed by NICER}

\author[0000-0001-5711-084X]{B. Rani}
\affiliation{NASA Goddard Space Flight Center, Greenbelt, MD 20771, USA}
\affiliation{Center for Space Science and Technology, University of Maryland Baltimore County, USA}
\affiliation{Korea Astronomy and Space science Institute, 776 Daedeokdae-ro, Yuseong-gu, Daejeon 30455, Korea}

\author{Jungeun Kim}
\affiliation{Korea Advanced Institute of Science and Technology, 291 Daehak-ro, Yuseong-gu, Daejeon 34141, Republic of Korea} 

\author{I. Papadakis}
\affiliation{Department of Physics and Institute of Theoretical and Computational Physics, University of Crete, 71003 Heraklion, Greece}
\affiliation{Institute of Astrophysics, FORTH, GR-71110 Heraklion, Greece}

\author{K. C. Gendreau}
\affiliation{NASA Goddard Space Flight Center, Greenbelt, MD 20771, USA}

\author{M. Masterson}
\affiliation{MIT Kavli Institute for Astrophysics and Space Research, Massachusetts Institute of Technology, Cambridge, MA 02139, USA}

\author{K. Hamaguchi}
\affiliation{NASA Goddard Space Flight Center, Greenbelt, MD 20771, USA}

\author{E. Kara}
\affiliation{MIT Kavli Institute for Astrophysics and Space Research, Massachusetts Institute of Technology, Cambridge, MA 02139, USA}

\author{S.-S. Lee}
\affiliation{Korea Astronomy and Space science Institute, 776 Daedeokdae-ro, Yuseong-gu, Daejeon 30455, Korea}

\author{R.~Mushotzky}
\affiliation{Department of Astronomy, University of Maryland, College Park, MD 20742, USA} 


\begin{abstract}
Variability studies offer a compelling glimpse into black hole dynamics, and NICER's (Neutron Star 
Interior Composition Explorer)
remarkable temporal resolution propels us even further.
NICER observations of an Active Galactic 
Nucleus (AGN), NGC 4051, have charted the geometry of the emission region of the central supermassive 
black hole. Our investigation of X-ray variability in NGC 4051 has detected extreme variations spanning 
a factor of 40 to 50 over a mere 10 to 12 hours. For the first time, we have constrained the X-ray Power 
Spectral Density (PSD) of the source to 0.1 Hz, corresponding to a temporal frequency of 
10$^4$~Hz in a 
galactic X-ray binary (GXRB) with a mass of 10~$M_{\odot}$. No extra high-frequency break/bend or 
any quasi-periodic oscillations are found. Through detailed analysis of energy-dependent PSDs, we found that 
 the PSD normalization, the high-frequency PSD slope as well as the bending frequency remains consistent 
across all energies within the 0.3-3 keV band, revealing the presence of a constant 
temperature corona. These significant findings impose critical constraints 
on current models of X-ray emission and variability in AGN.

\end{abstract}


\section{Introduction}
Active galactic nuclei (AGN) exhibit remarkable variability. The amplitude of 
variability increases and the time-scales of variability decrease as energy levels 
rise. The most rapid and largest amplitude variations in AGN are observed in X-rays 
(at least in radio-quiet AGN).
X--ray observations provide us a unique probe of the physical processes that 
operate in the immediate vicinity of supermassive black holes.  
Here, we present a comprehensive study of the fastest variability ever 
observed in NGC~4051, utilizing data from the Neutron Star Interior Composition 
Explorer (NICER). With its expansive effective area, NICER provides an exceptional 
temporal resolution of approximately 300 nanoseconds within the 0.2-10 keV energy 
range, enabling investigations into variability on timescales shorter than what 
has been possible before.

The advent of previous X-ray missions, notably the Rossi X-ray Timing Explorer 
(RXTE) and XMM-Newton, equipped with advanced 
observational capabilities, has yielded remarkable discoveries that have 
significantly advanced our comprehension of the 
temporal behavior, accretion processes, and physical characteristics of AGN. 
These investigations have unveiled noteworthy findings, including 
the detection of characteristic timescales, i.e.\ bending frequencies ($\nu_{bend}$),  
 quasi-periodic oscillations, 
and the correlation between X-ray variability  and the physical properties of the source  
\citep[e.g.][]{mchardy2004, markowitz2003, gierlenski2008, mchardy2006}. 
Additionally, 
the detection of Iron-$K_{\alpha}$ 
line features \citep{reynolds2009, fabian2005, nandra1997} and the manifestation 
of a soft X-ray excess in AGN X-ray 
spectra \citep{ghosh2022, filippo2008, done2006} indicate a compact corona 
and an accretion disk extending to the innermost stable orbit.

The characterization of X-ray variability has extensively utilized Power 
Spectrum Density (PSD) analysis. 
The PSD slopes and breaks are used to investigate their correlation with the 
underlying physical properties of AGN's central engines 
\citep[e.g.][and references therein]{mchardy2006, gonz2012}.
The PSD characteristics (slopes and breaks) have also been compared with 
black holes of lower masses, i.e., X-ray binaries (XRB). These studies favor 
that the dynamics of accretion are similar for different black hole 
masses \citep{gonz2012,  mchardy2006}, although the detailed 
scaling relations are not well determined.

The Seyfert galaxy, NGC 4051, has a redshift of 0.0023, and a black hole mass 
of 1.7$\times$10$^6$~M$_{\odot}$ \citep{denney2009}. 
Being one of the most variable AGN, the source has been extensively studied  
using various X-ray observatories 
\citep[e.g.][]{lawrence1987, papadakis1995, mchardy2004, vaughan2011}. 
Variations by a factor up to $\sim 6-7$ were  observed on time scales of the order of a 
few hundred seconds \citep{vaughan2011}. Its X-ray PSD, 
well-defined down to $\sim 10^{-2}$ Hz, shows a bend around 2$\times$$10^{-4}$~Hz 
\citep{vaughan2011, mchardy2004}. The 
PSD slope changes from $-$1.1 to $-$2.3 at the bend. Overall, the PSD shape of NGC~4051 is 
similar to soft state PSDs of XRBs \citep{mchardy2004}.

The NICER observations of NGC 4051 from 2017 to 2022 have been extensively examined 
in this study. The organization of the paper is as follows. Section 2 covers the 
methods - observations and data analysis. Section 3 is dedicated to PSD analysis, 
detailing the results. Summary and discussion  are given in Section 4.

\begin{figure*}
\includegraphics[scale=0.35, trim=0 0.2 0 0, clip]{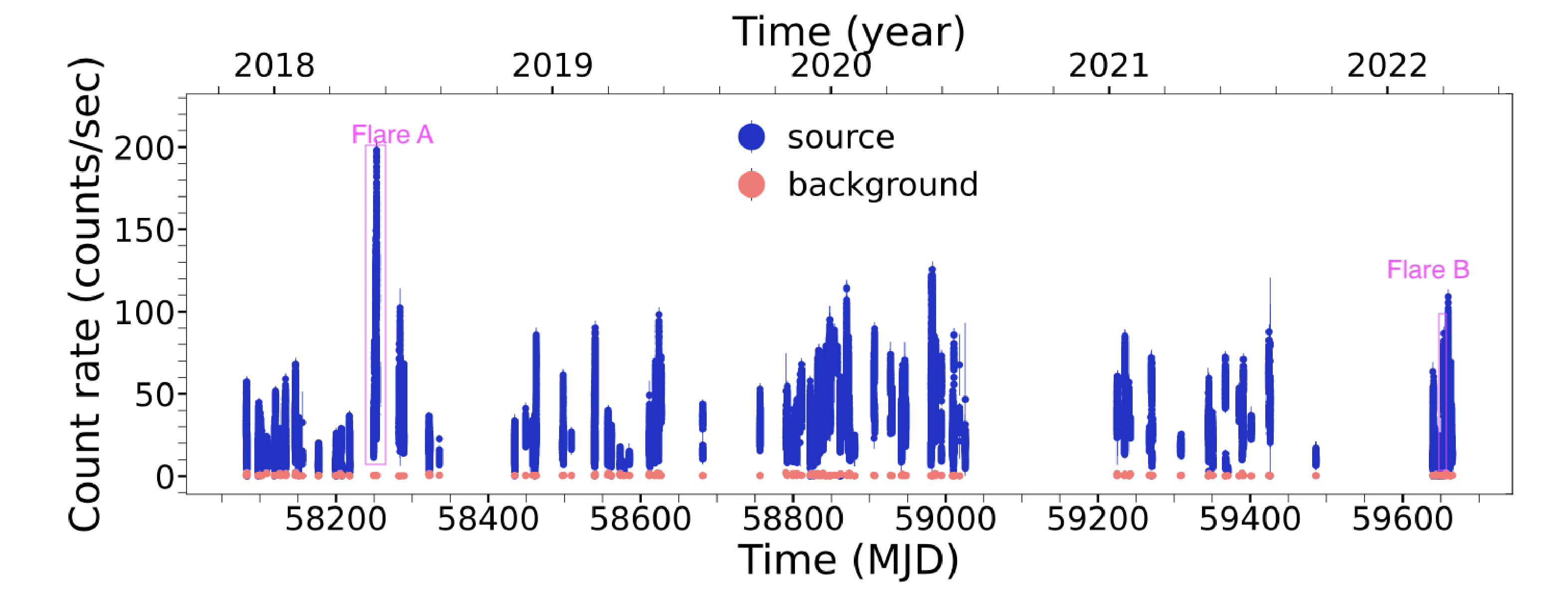}
\includegraphics[scale=0.48, angle =-90, trim=80 13 70 250, clip]{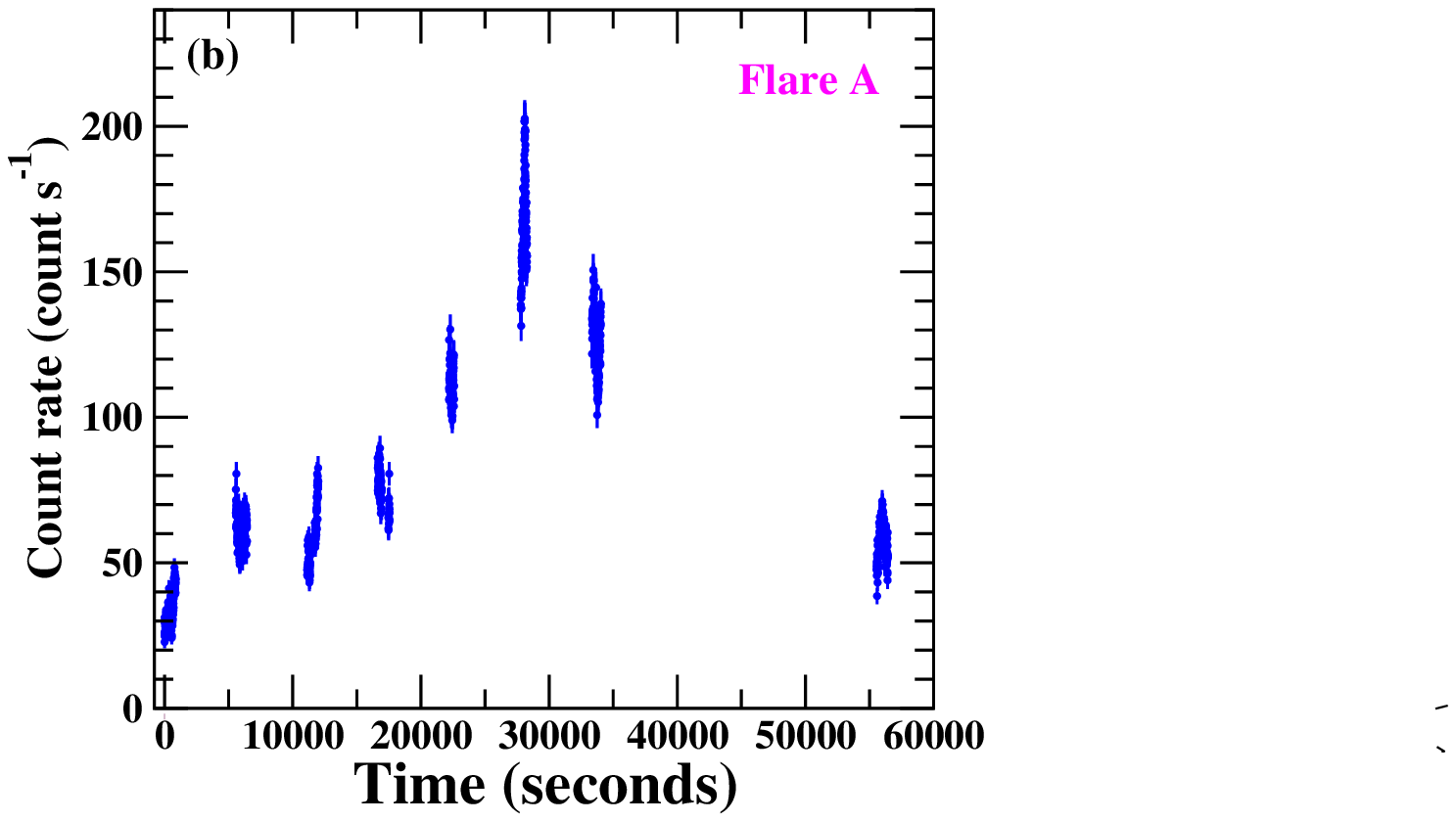}
\includegraphics[scale=0.48, angle=-90, trim=80 13 70 250, clip]{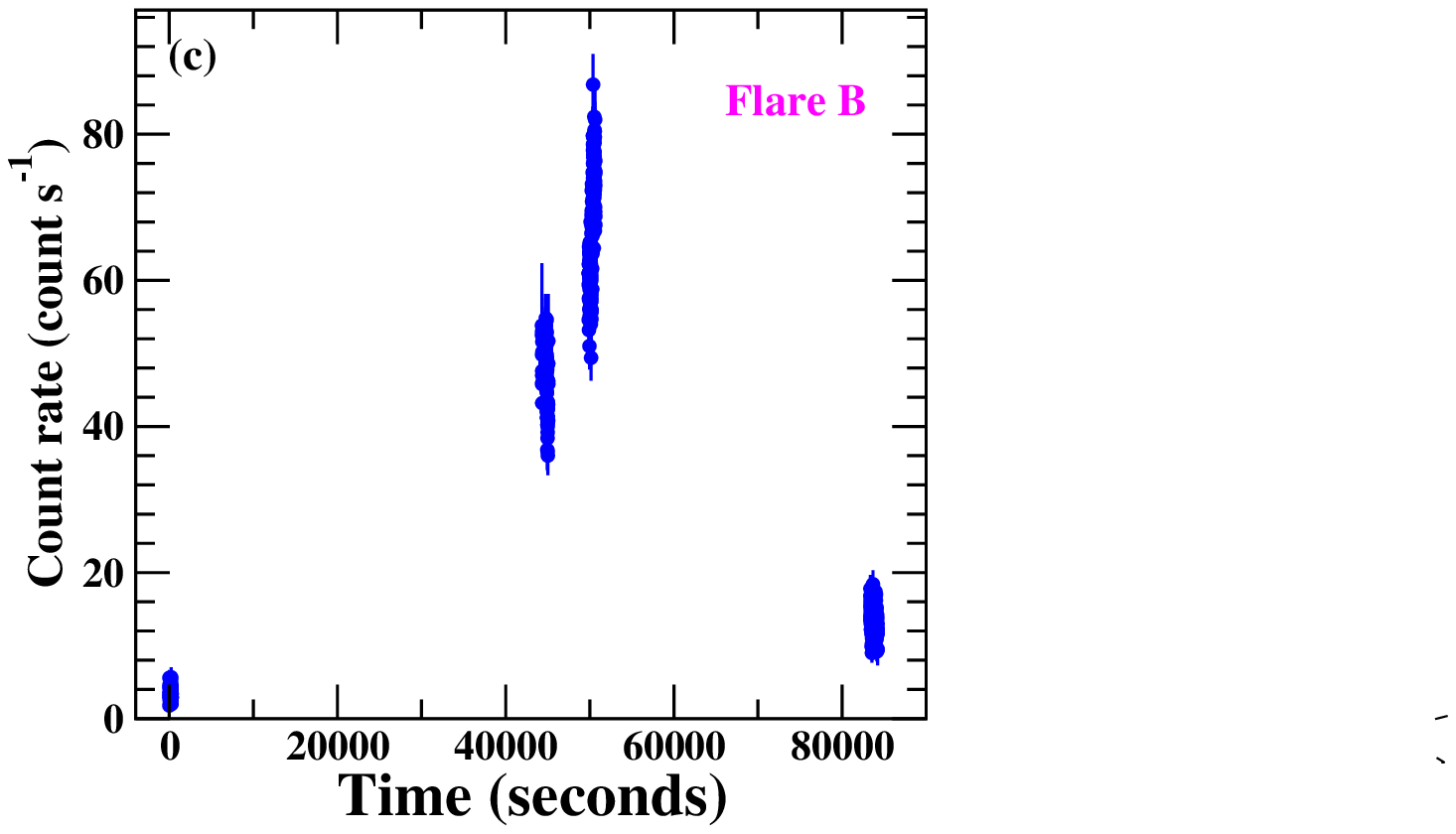}
\caption{{\bf Tracking down the swiftest variations in NGC~4051}: NICER light curve in the 0.3-3 keV energy range (in blue), extracted from archival 
observations taken between November 2017 and March 2022 ($\Delta t$ = 5~seconds).  The corresponding background 
count rate is plotted in red at the bottom.  {\it Bottom panels:} A zoomed version of the brightest, Flare~A (b), and the 
fastest flare, Flare~B (c)  (“0” corresponds to the start time of each event). }
\label{fig1}
\end{figure*}

\section{Methods - observations and data analysis}

\subsection{NICER observations}
With NICER, NGC~4051 has been observed many times between 2017 
and 2022 (see Fig.~1).  The archival {\it NICER} 
observations were reduced using $nicerl2$ 
pipeline\footnote{https://heasarc.gsfc.nasa.gov/docs/nicer/analysis\_threads/nicerl2/}. 
The data were 
processed using the NICER data-analysis software version 
2022-12-16$\_$V010a following standard calibration and filtering. 
The background contamination was filtered following the recommended step-by-step 
analysis\footnote{https://heasarc.gsfc.nasa.gov/docs/nicer/analysis\_threads/background/}. 
In addition to the standard $nicerl2$   
'overshoots' (0-500) and 'undershoot' (0-1.5) events filtering, events  30 degrees below the Earth limb and 
40 degrees below 
the bright Earth limb were flagged. 
To improve signal-to-noise ratio, we filtered noisy detectors (14, 34, 32, 41, 43, and 47). 
To extract the light curves in different energy bands, 
nicerl3-lc pipeline\footnote{https://heasarc.gsfc.nasa.gov/docs/nicer/analysis\_threads/nicerl3-lc/} 
was  used. The background light curve was constructed with the background estimator, 3C50 \citep{remillard2022} using the 
default setting. 
Given the low count rate of the source above 3~keV energies, we focused on the variability study 
in the 0.3-3.0~keV energy band.

Due to its higher count rate, NICER observations of the source can be binned with time intervals as short as 2~seconds. 
However, for consistent comparisons with XMM and across different energy bands, we use 5-second time bins.
A standard NICER observation comprises uniformly sampled data 
segments/chunks with a duration ranging from approximately 200 to 2000 seconds. 
Equal length 
evenly sampled data segments  are selected for the PSD analysis. For the PSD 
analysis light curves are sampled  in the following 
energy bands: 0.3-3.0, 0.3-0.5, 0.5-0.7, 0.7-1.0, and 1-3 keV (see section \ref{psd_analysis} for details). 

\subsection{NICER: probing rapid variations}

Figure \ref{fig1} shows the 0.3-3~keV source (in blue) and background (in red) light curve of NGC~4051.  
Prominent flux variations are evident in the long-term  light curve of the source. 
The source is variable at all sampled timescales, and variations by a factor of up to 
8–10 within a few kiloseconds are common. 
To search for rapid variability events, for intensity variations from $t_i$ to $t_f$, we establish a 
threshold defined as
\begin{equation}
 \frac{I(t_f = t_i +\Delta t_{flare})}{I(t_i)} \geq 20. 
\end{equation}
Intensity variations exceeding a factor of 20 within a timescale  of 1~day, i.e.\ $\Delta t_{flare} \leq 1~day$  
are classified as extreme variability events.
Over a span of 
approximately 4 years of NICER observations, we recorded 20 extreme variability events. 
Two such extreme flares  are highlighted in 
boxes in Figure 1, and a zoomed version is presented in the bottom panels. Figure 1(b) showcases the brightest 
flare (Flare~A) recorded for the source. The source intensity changed by a factor of 10 (count rate varying 
from $\sim$20 to $\sim$202) within 28 kiloseconds (7.8 hours). 
Flare B, plotted in panel (c) corresponds to an  X-ray intensity variation by a 
{\bf factor of 50} (count rate ranging from 1.8 to 87) recorded within 45 kiloseconds (12.4 hours).

\subsection{XMM-Newton observations} 
We analysed all the available XMM-Newton observations of NGC~4051 taken between 2001-2018 (see Table \ref{table_Xmm}). Note that 
most of the 
observations were conducted in 2009. The extraction of light curves 
was carried out using 
the interactive analysis available in the XMM-Newton 
archive\footnote{http://nxsa.esac.esa.int/nxsa-web/}.
For each observation, source events were extracted using a 40-arcsec circular region centered 
on the target, while background events were extracted from a nearby 40-arcsec circular region 
that did not overlap with the source region. The source background is found to be relatively low 
(count rate $\leq$2 counts/sec, average count rate $\sim$0.3) and mostly stable throughout 
the observations. Light curves were extracted from the EPIC pn data in the same energy bands and 
same time binning ($\Delta$$t$=5~sec) as those 
used in the NICER observations.

\section{Results - Power-Spectral density analysis} 
\label{psd_analysis}
We investigated  the variability characteristics of the source by employing 
the PSD analysis method. 
For the NICER data, we computed PSDs using segments of uniformly sampled data, 
each spanning a duration of 1000 seconds. Our dataset consisted of 104 such intervals. 
We computed periodograms of each one following \cite{vaughan2011}, and subsequently we 
computed their averages. 
The periodogram normalization unit is $(\text{rms}/\text{mean})^2 \, \text{Hz}^{-1}$ or $\text{Hz}^{-1}$.
For the XMM dataset, we concentrated on uninterrupted data intervals lasting 20 
kiloseconds, resulting in the identification of a total of 24 such intervals. Given 
that the number of periodograms over all observations was less than 50, we adopted the 
strategy of averaging and applying binning with 2  points\footnote{We need 50 
periodograms when computing the mean PSD to ensure Gaussian statistics \citep[see][for 
further details]{papadakis1993}.}.

Figure 2 shows the PSDs in the  0.3-3~keV energy range for the XMM (in red) and 
NICER data (in blue).  
A striking resemblance of the PSD shapes for the overlapping frequencies ($\geq$0.001~Hz) can be 
seen here. 
Moreover, NICER observations  allow us to decrease the Poisson 
noise level by a factor of  $\sim$3 (see Fig.~2), allowing us to better 
constraint the PSD shape up to $\sim 0.1$ Hz (see Fig.~3). 
 This corresponds to a frequency of $10^4$ Hz for a ten solar mass black hole in an 
X--ray binary (assuming a black hole mass of $\sim 10^6$ M$_{\odot}$ for NGC 4051 and that 
 timescales are proportional to $M_{BH}$). To 
the best of our knowledge, variability at such 
high frequencies have never been studied before, for any accreting black hole.

\begin{figure}
\center
\includegraphics[scale=0.37, angle =-90, trim=0 33 50 0, clip]{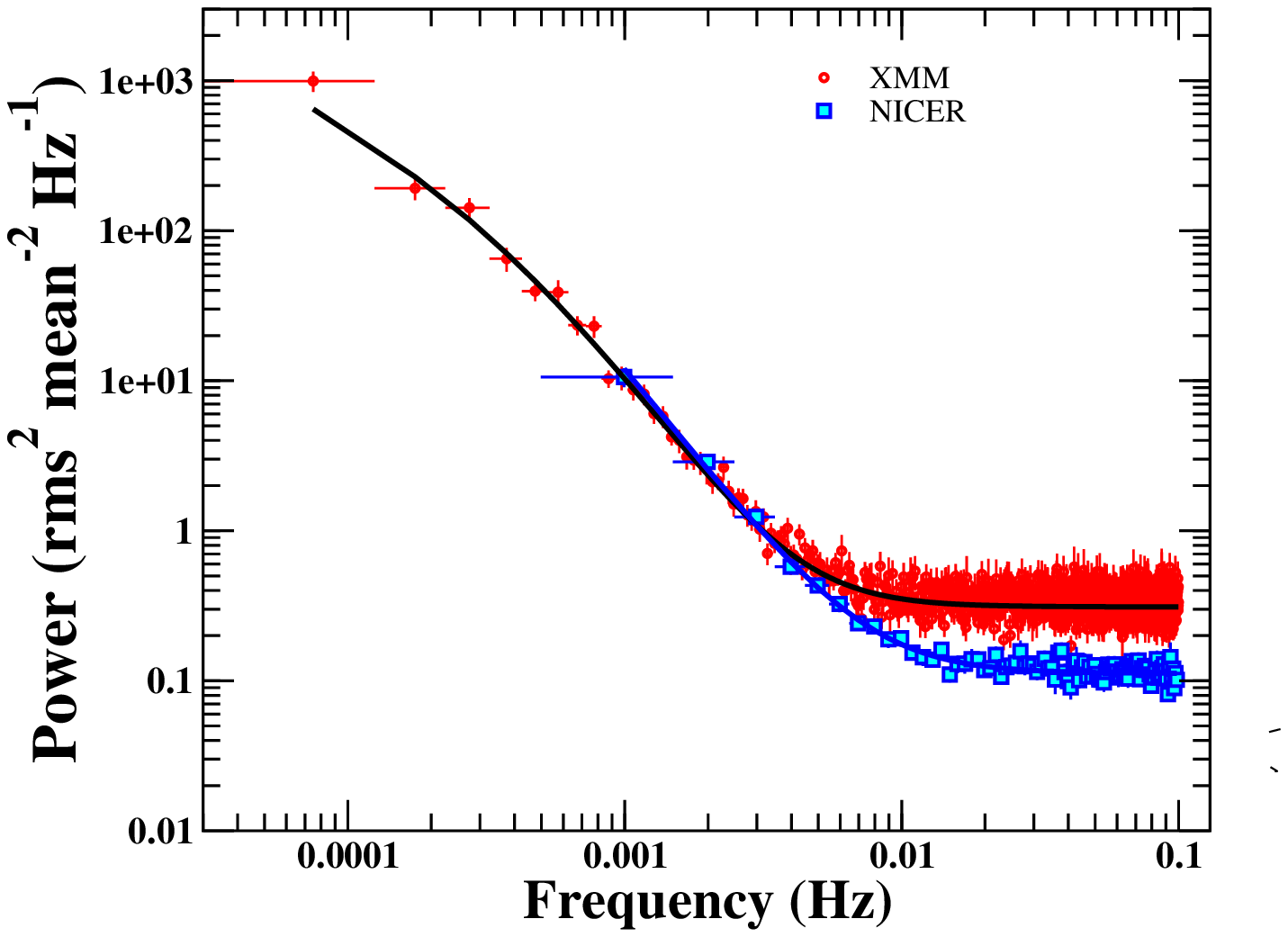}
\includegraphics[scale=0.34, angle=-90, trim=50 0.2 0 0, clip]{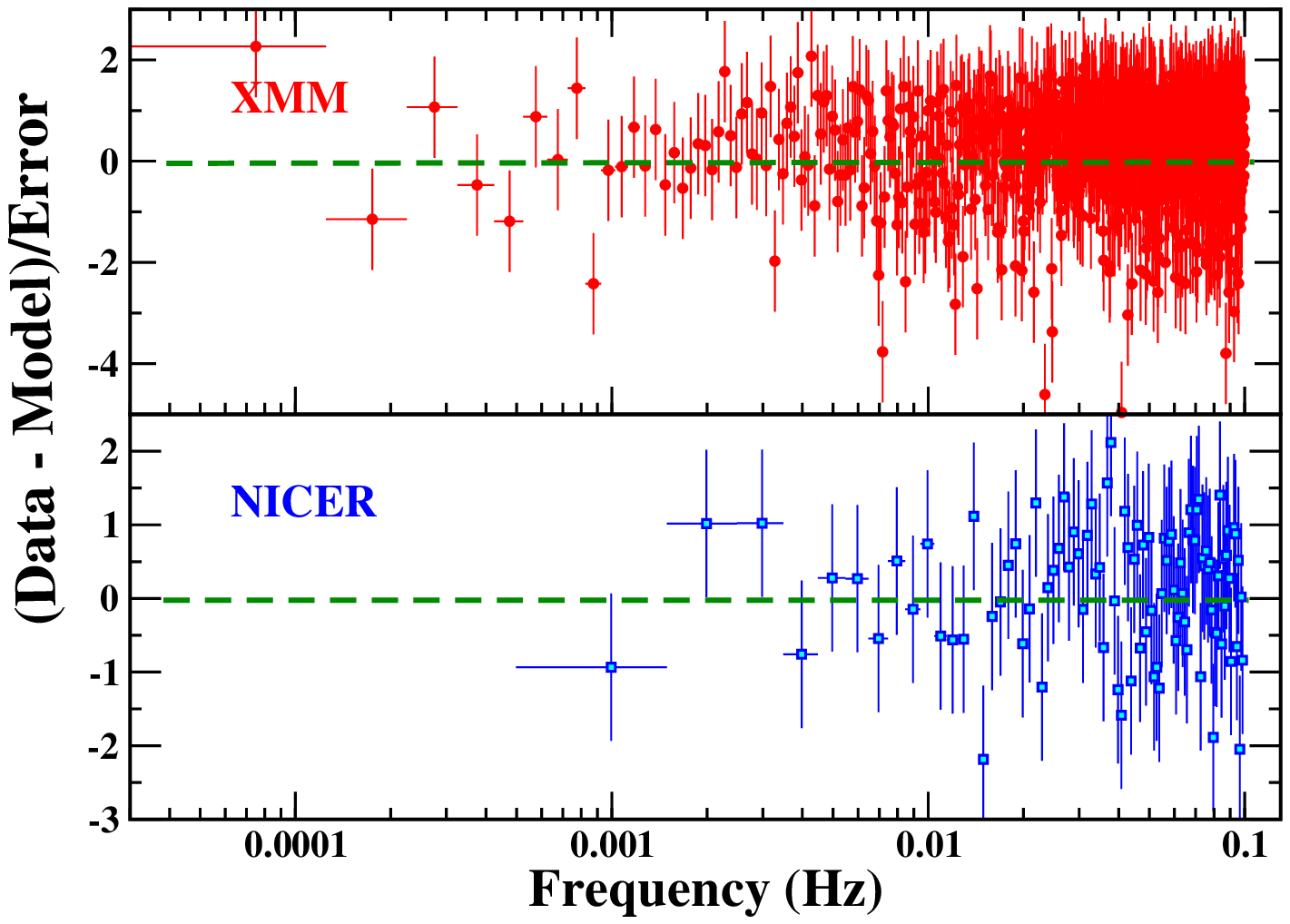}
\caption{{\bf Power-spectral density analysis:} {\it Top:} PSDs derived from the 0.3-3.0~keV light curves from 
XMM-Newton (red circles) and NICER (blue squares). The black and magenta lines represent the 
best-fit models for the XMM-Newton  and NICER PSDs, respectively.
{\it Bottom:} Best-fit residuals for the XMM and NICER PSDs.   }
\label{fig2}
\end{figure}

\begin{figure}
\center
\includegraphics[scale=0.37, angle =-90, trim=0 33 30 0, clip]{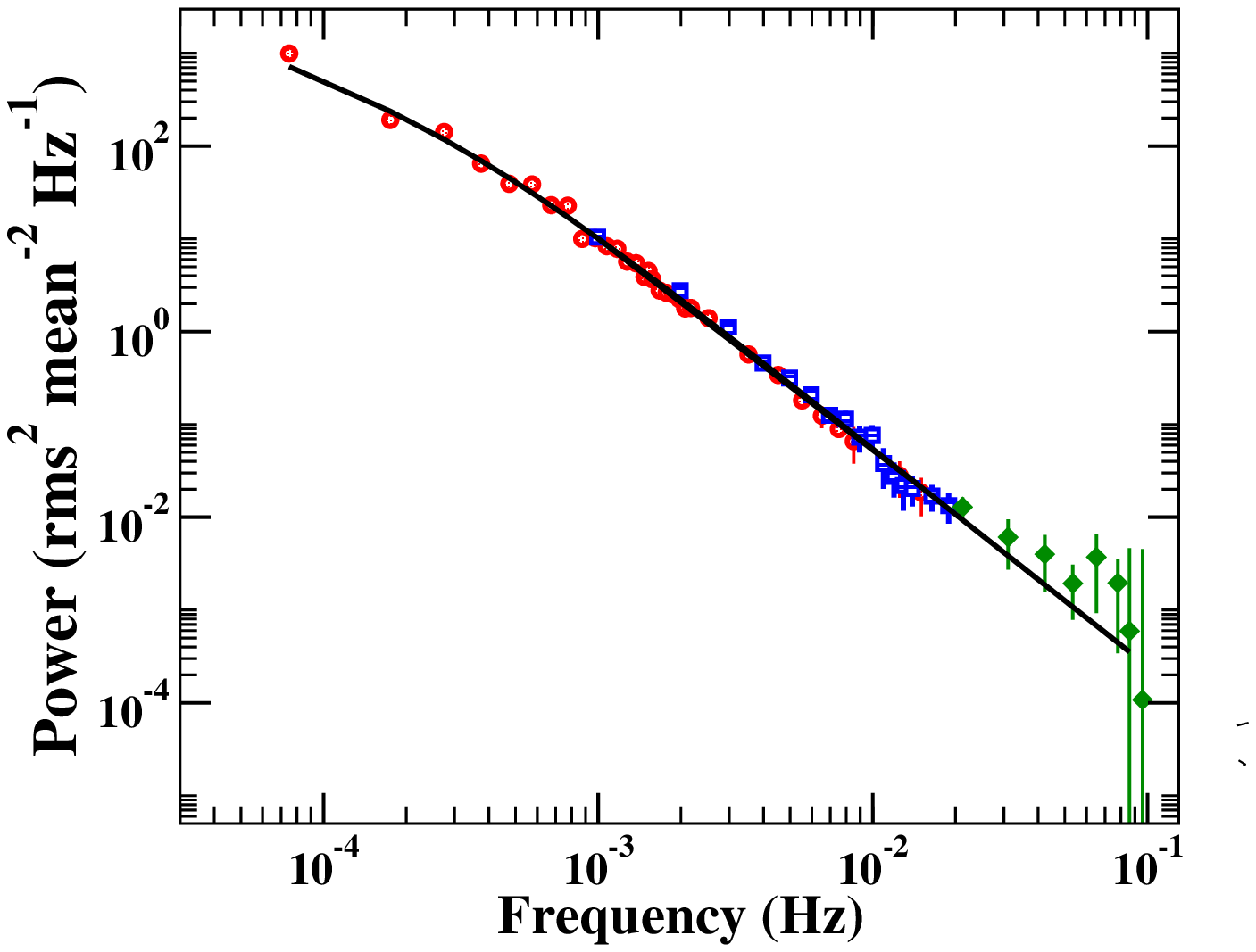}
\caption{{\bf Exploring uncharted variability at high frequencies:} Poison noise subtracted PSDs derived from light curves 
obtained from NICER  and XMM-Newton. Heavily binned PSD points are in green. }
\label{fig3}
\end{figure}

 \subsection{Model fitting and stationarity of the PSD shape}
At first, we individually fitted the full band (0.3-3.0~keV) PSDs from XMM-Newton and NICER.
As has been noted via pervious studies 
\citep{mchardy2004, vaughan2011}, 
the X-ray PSD of the source follows a broken/bending power-law (BPL), with a break 
frequency $\sim$2$\times$10$^{-4}$~Hz. 
To 
confirm that the XMM PSD does require a bending frequency, we first fitted it with a 
simple power-law (PL), defined as 
\begin{equation}
PSD_{PL}(\nu) = N (\frac{\nu}{10^{-3}})^{-\alpha} + Pnoise, 
\end{equation}
where $N$ is the PSD normalization i.e.\ the PSD value at $\nu = 10^{-3}$~Hz, 
$\alpha$ is the slope, and 
$Pnoise$ is the constant power component that appears at high frequencies due to the Poisson noise. The best-fit model resulted in $\chi^2_{\rm min}$ = 200 with 146 dof. Then, we used a BPL model to fit the XMM PSD, as follows:  
\begin{equation}
\label{eqn:bend}
  PSD_{BPL}(\nu) = \frac{N \nu^{\alpha_{\rm low}}}{  1 + ( \nu / \nu_{\rm bend} )^{\alpha_{\rm low} - \alpha_{\rm high} } }  + Pnoise ,
\end{equation}
where $N$ is the normalization, and $\alpha_{\rm low}$, $\alpha_{\rm high}$ are the  slopes below  and   
above $\nu_{bend}$.
In fitting the PSD, we fixed the lower frequency index at $\alpha_{\rm low} = -1$ \citep[as it is much better 
constrained by the RXTE data;][]{mchardy2004}. 
The BPL model provided a better fit
with $\chi^2_{\rm min}$ = 153 with 145 dof, resulting in $\Delta \chi^2$ = 47 for an extra degree of freedom (compared to the PL best fit).  This result strongly favours the BPL model over the
PL model. However, NICER observations cover only a frequency range up to $10^{-3}$ Hz, which is above the 
bending frequency that is detected in the XMM PSD. Indeed, the PL model fits the NICER PSD very 
well ($\chi^2_{\rm min}=17$ for 26 dof.)

\begin{table*}
\center
\caption{Best-fit results to the XMM–Newton and NICER PSDs}
\label{table:bestfitstoxmmandnicer}
\begin{tabular}{ | l c c c c c  | } 
  \hline
Energy band & Normalization         &  $\alpha_{high}$/$\alpha$   & $\nu_{bend}$                     &P$_{noise}$                             & $\chi^2$/dof \\
(0.3-3.0~keV)       & (A)                   &                            &  $\times 10^{-4}$     (Hz)                       &                               &       \\\hline
XMM-Newton     & 0.04$^{-0.01}_{+0.02}$ & $-$2.47$^{-0.11}_{+0.10}$  & 4.2$^{+1.6}_{-1.3}$ & 0.30    & 153/145 \\   
NICER          & 14.03$^{-3.06}_{+3.65}$  & $-$2.38$^{+0.15}_{-0.15}$ & - & 0.11 & 17/26 \\
XMM-Newton + NICER & 0.05$^{-0.01}_{+0.01}$ & $-$2.38$^{+0.08}_{-0.08}$ & 3.7$^{-1.2}_{+1.4}$ &0.30/0.11     & 183.8/173      \\\hline 
\end{tabular}\\
Note: The errors on the best-fit parameters correspond to the 90$\%$ confidence interval. \\
\end{table*}

\begin{table*}
\center
\caption{Best BPL model fits to combined XMM–Newton and NICER PSDs in different energy bands}
\begin{tabular}{ | l c c c c c c c | } \hline
Energy  & Norm                 &$\nu_{bend}$  & $\alpha_{high}$         & $\alpha_{high}$                       &  $\alpha_{high}$          & P$_{noise}$ & $\chi^2$/dof      \\
 (keV) & (A)                   &$\times 10^{-4}$(Hz)  &  XMM                    & NICER                         & tied                      & XMM/NICER   &                    \\ \hline 
0.3-0.5     & 0.05$^{+0.02}_{-0.01}$&3.9$^{+1.6}_{-1.3}$  &$-$2.43$^{+0.14}_{-0.16}$&$-$2.29$^{+0.09}_{-0.10}$  &                           & 0.90/0.34   & 157/176                \\ 
     & 0.04$^{+0.01}_{-0.01}$&3.3$^{+1.8}_{-1.5}$  &                         &                           &$-$2.34$^{-0.12}_{+0.11}$  &    & 168/176       \\ 
0.5-0.7     & 0.05$^{+0.03}_{-0.02}$&3.5$^{+1.6}_{-1.4}$  &$-$2.38$^{+0.15}_{-0.17}$&$-$2.23$^{+0.11}_{-0.10}$  &                           & 1.33/0.50   & 153/176      \\  
            & 0.06$^{+0.03}_{-0.02}$&3.1$^{+1.6}_{-1.4}$  &                         &                           & $-$2.31$^{+0.13}_{-0.14}$ &    & 170/176      \\  
0.7-1.0     & 0.05$^{+0.04}_{-0.02}$&3.4$^{+2.1}_{_1.7}$ &$-$2.35$^{+0.18}_{-0.20}$&$-$2.22$^{+0.12}_{-0.13}$  &                           & 1.98/0.67   & 191/176          \\
            & 0.06$^{+0.05}_{-0.02}$&2.6$^{+2.0}_{-1.5}$ &                         &                            &$-$2.22$^{+0.14}_{-0.15}$  &    & 196/176      \\
1.0-3.0     & 0.06$^{+0.02}_{-0.01}$&2.7$^{+1.1}_{-1.2}$  &$-$2.25$^{+0.12}_{-0.14}$&$-$2.28$^{+0.11}_{-0.13}$  &                           & 1.39/0.41   & 172/176     \\ 
            & 0.06$^{+0.03}_{-0.02}$&2.9$^{+1.6}_{-1.2}$  &                         &                            &$-$2.28$^{+0.12}_{-0.11}$  &    & 172/176    \\\hline 
\end{tabular}
\end{table*}
 
The best-fit results are listed in Table \ref{table:bestfitstoxmmandnicer}. The black and blue solid lines in Fig.~2 indicate the best-fit models to the XMM and NICER PSDs, respectivly, while the lower panels in the same figure show the best-fit residuals. In all cases, we fit the PSDs by keeping the Poisson noise level fixed at a value that we determine by fitting a straight line
above a frequency of 0.03 Hz for the XMM PSD and 0.04 Hz for the NICER PSD. 
Furthermore, we considered PSD points at frequencies lower than 0.015~Hz only 
while fitting the models to the data. In this way, 
the fit is determined by the low-frequency part of the PSD and not by the higher frequencies, where there are
more PSD estimates, but they are all flat due to the Poisson noise component domination.
The XMM PSD is well described by the BPL model defined by eq.~3, with  
$\nu_{\rm bend}$ = 4.2$^{-1.3}_{+1.6}\times 10^{-4}$  
and   $\alpha_{high}$ = $-$2.47$^{-0.11}_{+0.10}$. 
The PL model (eq.~2)  described well the NICER PSD, with a PSD slope, 
$\alpha$, of $-$2.38$^{-0.15}_{+0.15}$.  

\subsection{PSD tests of stationarity}
The high-frequency slopes obtained from the model fits to the two PSDs are consistent with each 
other within the error bars.  This result implies that the XMM and the NICER PSDs are similar 
in shape, in agreement with the visual inspection of the two PSDs in Fig.~2.  This agreement 
indicates that the XMM-Newton PSD, primarily based on observations from 2009, aligns closely 
with the NICER PSD, derived from observations that span 2017 to 2022, despite a significant 9-year gap. 
To investigate this issue further, we simultaneously fitted the XMM and the NICER PSDs together 
with the BPL model. The best-fit results are also listed in Table \ref{table:bestfitstoxmmandnicer}. 
The fit is very good, with $\chi^2_{\rm min}=183.8$ for 173 dof ($p_{\rm null}=0.27$). The bending 
frequency, $\nu_{\rm bend}$, is constrained to be $3.7^{-1.2}_{+1.4}\times 10^{-4}$ Hz. 
Furthermore, our fit returns $\alpha_{high} = -2.38^{-0.08}_{+0.08}$, which is remarkably 
similar to the best fit values obtained from the fits to the XMM and NICER PSDs. 

When we fit the two PSDs individually with the BPL and the PL models, the combined $\chi^2_{\rm min}$ is 
170 for 171 dof. Therefore, $\Delta\chi^2=13.8$ for two extra dof when we fit the PSDs simultaneously. 
Statistically speaking, this increase in $\chi^2_{\rm min}$ is significant 
($p_{\rm null}=1.3\times 10^{-3}$), therefore the fit to the individual PSDs appears to be a 
better one. The biggest difference between the XMM and the NICER PSDs is their amplitude at low 
frequencies, while the PSD slopes are almost identical. For example, the best-fit BPL model for 
the XMM PSD predicts that PSD ($\nu=10^{-3}$ Hz) $\sim 6.1$, which is smaller than the best PL fit 
to the NICER data at the same frequency (see Table~1). We suspect that this difference is due to 
red-noise leakage. Since the lowest frequency in the NICER spectrum is higher than 
$\nu_{\rm bend}$ and $\alpha_{\rm high}$ is steeper than 2, we expect the NICER PSD to be 
affected by red noise leakage, hence the slightly higher nornmalization of the NICER PSD. 
The excellent BPL fit and agreement in PSD slopes (within errors) suggest the two PSDs are 
consistent.

Our results imply that there is no strong evidence of time evolution of the power spectrum over 
time scales of the order of a few years. The current light curves do not allow us to examine where 
the PSDs remain the same on shorter time scales, within the time periods of the NICER or XMM observations. 
As a step towards such an investigation, we computed the PSD during the extreme flaring and the 
non-flaring parts of the NICER observations and fitted the resulting PSDs with the PL model defined 
by eq.~2. Our results show that the PSDs are almost identical. The best-fit PSD sloes 
are $-2.34\pm 0.21$ and $-2.24\pm 0.11$ for the flaring and non-flaring parts, while the respective 
PSD normalizations are 19.2$^{+7.3}_{-7.3}$ and $14.7_{-2.2}^{+2.4}$. These results suggest that 
there may be no major differences between the PSDs on shorter time scales as well. 

\subsection{Extra components in the high frequency PSD of NGC 4051}
As a next step, we explored the potential presence of an additional high-frequency break in the PSDs. 
First, we fitted the XMM-Newton and NICER data with a model that includes a second frequency 
break, and we limited the best-fit second bending frequency to be higher than the break frequency 
at $\sim 4\times 10^{-4}$ Hz which is already detected. The best-fit results indicated that there 
were no additional high-frequency breaks. 

Figure 3 shows the Poisson noise-subtracted XMM and NICER PSDs (red and blue poits, respectively). We 
have heavily binned both PSDs (together) above 0.015 sec$^{-1}$  (green points; the bins consist of 
200 points in this case). 
The black solid line shows the best BPL fit to the XMM and NICER data. This plot clearly shows 
that the PSD extends up to a frequency of 0.1 Hz with the same slope, and there is no
indication of a further break at higher frequencies, up to 0.1 Hz. 
Finally, despite the high signal-to-noise ratio and the broad frequency coverage, Fig.~3 also 
shows that there is no evidence of 
a quasi-periodic oscillation over the sampled frequency range in the X-ray PSD of NGC~4051. 

\begin{figure}
\center
\includegraphics[scale=0.35, angle =-90, trim=80 3 5 0, clip]{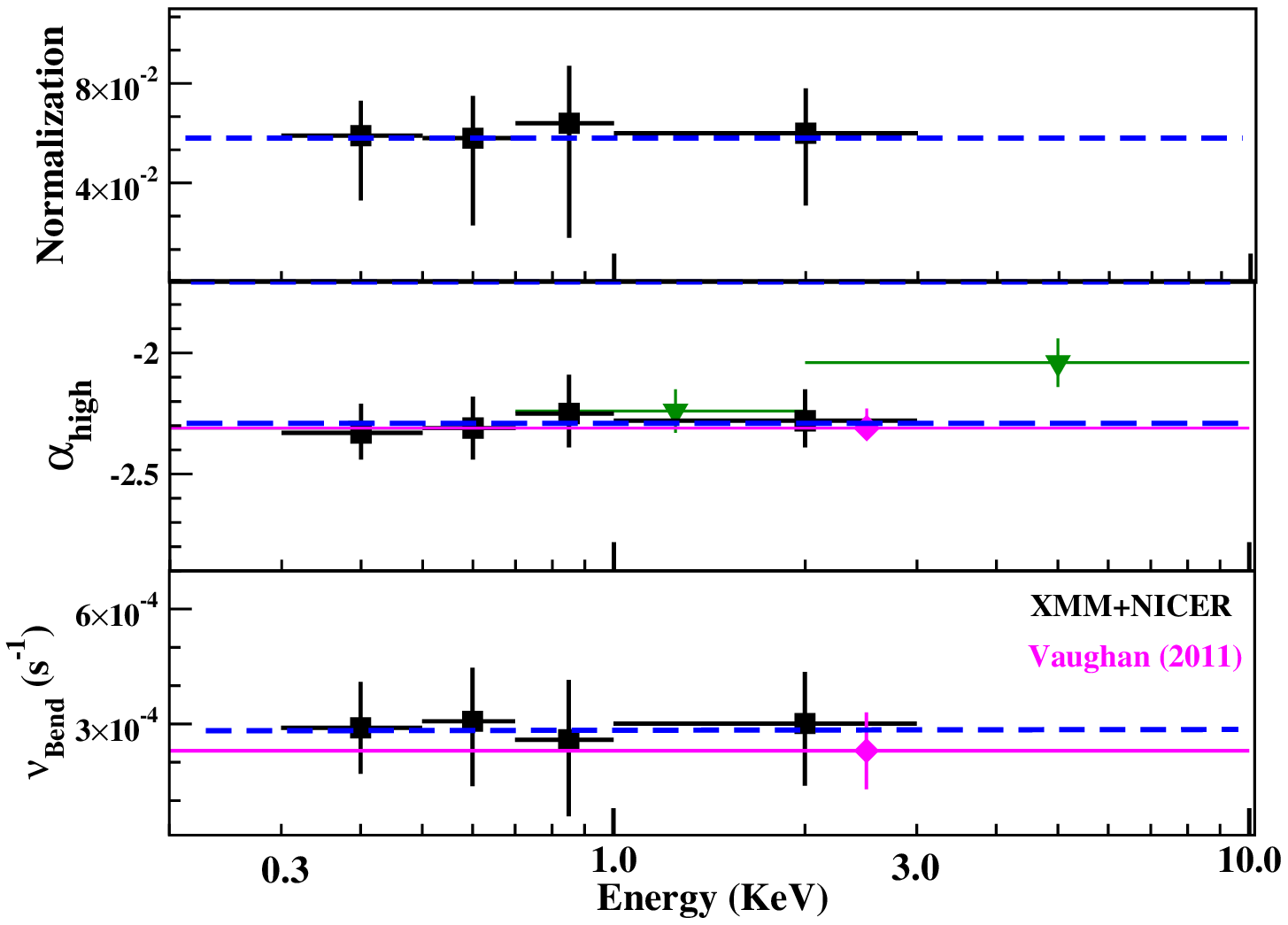}
\caption{ Best-fit PSD parameters, i.e., normalization, high-frequency PSD slope, and bending frequency 
(from top to bottom), plotted as a function of energy. The weighted mean of the PSD parameters is marked 
by the blue dashed lines. }
\label{fig4}
\end{figure}

\subsection{Energy dependence of the PSD shapes}
The final course of action is to investigate the energy dependence of the PSD parameters, i.e.\ PSD amplitude, 
$\nu_{bend}$, and $\alpha_{high}$. To achieve this, we split the light curves 
into different energy bands and computed PSDs (see Figs.~5-8), as discussed in Section 3.2. 
The BPL model (eq.~2) is used to fit the XMM and NICER PSDs together. Apart from the 
high-frequency slope ($\alpha_{high}$), the other model parameters are tied 
during fitting. The best-fit 
parameters are listed in Table 2. Figures 5-8 illustrate the PSDs and best-fit models 
with $\alpha_{high}$  fitted separately. 
In comparison to the XMM PSDs, NICER PSDs 
have a slightly 
flatter $\alpha_{\text{high}}$ below 1~keV, but they are in agreement within the uncertainties. 

We also investigated whether fitting the same $\alpha_{\text{high}}$ (tied) to the two 
PSDs yields a good
fit. The best-fit parameters are listed in Table 2. We found that this fitting approach 
provides a good 
fit and leads to reduced uncertainty for the $\alpha_{\text{high}}$ and $\nu_{\text{bend}}$ 
parameters. We used the parameter 
values obtained from this approach to investigate the energy dependence of the PSD parameters.
  
Figure 4 presents the best-fit parameters - PSD normalization, $\nu_{\text{bend}}$, 
and $\alpha_{\text{high}}$ — plotted as a function 
of energy. In addition to the results from this study, we have included the parameters from \cite{vaughan2011} 
(magenta diamonds), extending the energy range up to 10 keV. Our analysis shows that the PSD normalization, 
 $\nu_{\text{bend}}$, and $\alpha_{\text{high}}$   remain unchanged across different energies. Specifically, $\nu_{\text{bend}}$ from our study 
aligns with the values reported by \cite{vaughan2011} and remains consistent up to 10 keV. The PSD slope, $\alpha_{\text{high}}$, 
also remains consistent based on our data (with the dashed blue line representing the median) and is similar to  
$\alpha_{\text{high}}$ measured by \cite{vaughan2011} at low energies.
 However, a flattening of $\alpha_{\text{high}}$ above 2~keV had been found by 
\cite{vaughan2011} (see green triangles in Fig.~4). 

\section{Summary and Discussion}
With its exceptional temporal resolution, NICER has probed temporal characteristics of AGN variability 
on minuscule timescales that were previously beyond the reach of X-ray astronomy. This study covers 4 
years (2017-2022) of NICER observations of NGC 4051, during which the source intensity rate reached up to 200.  
We recorded 20 extreme flaring events in the source, with 
variability changing by a factor of $>$20 within a day. In the fastest recorded event, the source intensity 
changed by a factor of 50 in about 12 hours. Since NGC 4051 has a relatively small black hole mass ($10^6 M_{\odot}$), 
AGN with larger masses will probably show variability with smaller amplitudes over similar timescales. 
Nonetheless, variations by a factor of 10-20 within a 
few hours are probably common in AGN with small BHs at X-rays.

NICER observations allowed us to study the PSD in NGC 4051 at frequencies up to 0.1~Hz. 
Never before has variability at such high frequencies been explored in any accreting compact object. The 0.3-3 keV PSD, over three decades of frequencies above $10^{-4}$~Hz, is well fitted by a 
bending power-law with $\alpha_{high} = -2.38$ and $\nu_{\rm bend}$ = 3.7$\times 10^{-4}$~Hz. We do not detect any QPOs and/or any further slope break frequencies up to $\sim 0.1$ Hz. 
In addition, the PSD shape and normalisation have remained consistent over at least 10 years (this difference is noted between the 2009 XMM observations, which dominate the calculations of the XMM PSDs, and the middle of the NICER light
curves). This timescale corresponds to a period of approximately 3100 seconds for a 10 $M_{\odot}$ black hole in a GXRB. 
In GXRBs, the power spectrum typically remains constant during the individual states the system experiences throughout its evolution. GXRBs spend much more than 3000 seconds in each state. Thus, the stationarity of NGC 4051's X-ray PSD is not surprising and further indicates that the X-ray variability properties are similar between AGN and GXRBs.

The results for NGC 4051 differ from those for AGN IRAS 13224-3809 \citep{alston2019}. While 
IRAS 13224-3809, likely in the very high/intermediate GXRB state, shows increased PSD 
normalisation with decreasing flux, NGC 4051’s PSDs are flux-independent. Moreover, IRAS 
13224-3809’s PSD has two breaks, whereas NGC 4051’s extends to low frequencies without 
an extra break \citep{mchardy2004}. Despite these differences, both exhibit a constant 
PSD break (from -1 to a steeper slope) across energy bands, including 3–10 keV. Studies on more 
AGN are required to investigate further the issues of stationarity and the 
energy dependence of the PSDs in AGN.

Leveraging the high count rate of NICER observations, we delved into an energy-dependent PSD analysis covering 
almost a decade in energy width, ranging from 0.3 to 3.0 keV. This exploration provided a comprehensive view  of the system dynamics.
Our findings are striking: despite the large energy range, both the PSD normalization and 
bending energy remain constant at all energies, with $PSD_{\rm norm}=0.055\pm 0.007$, and $\nu_{bend}=(3.0\pm 0.4)\times 10^{-4}$. 
Moreover, the high-frequency PSD slope also remained 
constant within this energy band, and then it may flatten by $\Delta \alpha_{high}$ $\simeq$ 0.25 at energies 
higher than 2-3 keV, as it appears when we compare our results with those reported by \cite{vaughan2011}) The 
constancy of the PSD characteristics with energy is one of the most important results of our work, and should  place 
interesting constraints (discussed below) to all models for the X-ray emission and variability in AGN. 

\subsection{Implications for the origin of the soft-excess}
A soft excess component is clearly visible in the X-ray spectrum of NGC 4051 at energies below 
$\sim 1$ keV \citep{pounds2004, lobban2011}. It is possible that it can contribute up to $\sim 50$ per 
cent of the observed flux at energies below $\sim 0.5$ keV. 
The soft excess could be due to thermal comptonization of the disc photons in a warm (KT $\sim$ 0.1-1 keV) 
and optically thick ($\tau$ $\sim$ 10-20) corona, which may be responsible for the UV to soft 
X-ray emission in AGN \citep{petrucci2013}. According to this model, the warm corona could be located 
on top of the accretion disk at  radii which are larger than 10-20~$R_g$ \citep[see Fig.~10 in][]{petrucci2013}. 
In this model, the hot corona, which is responsible for the X-ray continuum emission at energies higher 
than $\sim 1-2$ keV, is located at even smaller radii, closer to the BH. 

If the soft excess is due to emission from the warm corona, then it is difficult to understand why the 0.3-0.5 and the 
1-3 keV PSDs are almost identical in NGC 4051, given the fact that the warm and the hot corona should be at 
different locations, and their physical properties are very different (optical depth of $\sim 0.5-1$ and $10-20$ for the 
hot and warm corona, respectively, while their temperature may be different by a factor up to a 100). It seems much more probable then 
that all observed variability, in all energy bands, originates from the hot corona. 

In this case, the warm corona could respond to and / or reprocesses the variability of the hot corona, and therefore contributes a component to the power spectrum with the same shape. It could also be that the warm corona does not vary in the frequency range that is analysed here and its
emission should be stable over the time scales we sample with our PSD analysis. However, in either case,
we would expect the PSD normalisation in the  0.3-0.5~keV band to be significantly smaller. For example, 
if the variability of the warm corona on short timescales is constant, we would expect the PSD 
normalisation of 0.3-0.5 lkeV to be $\sim$ 4 times smaller than the PSD normalisation in the 
1-3~keV band, assuming that the intrinsic PSD normalisation (of the hot corona) does not 
depend on energy, and
the flux contribution of the warm corona in the 0.3-0.5 keV is comparable to the contribution of the 
hot corona in the same band. Despite this prediction, we found identical PSD normalisation in the 
two bands. Therefore, our findings refute the hypothesis that a warm corona is behind the soft excess, 
regardless of its variability. In a more general note, our study also contradicts the idea of a 
constant soft excess, as the PSD normalization is independent of energy.

Another possibility is that the soft excess is a consequence of X-ray illumination of the accretion disc \citep{crummy2006}. 
It is not easy to predict the resulting PSD in this case since both the primary emission from the corona and the 
X-ray reflection component appear in each band. \cite{papadakis2016} have studied the expected PSDs in this case, 
but focusing on the 5--7 keV band. Figure 2 in that paper shows a ratio of the observed over the intrinsic X-ray PSD 
in the case of X-ray reflection in the lamp-post geometry. If the situation is similar at lower energy bands as well, 
then we may expect the PSDs in the 0.3-0.5 keV band (mainly) to show some "wiggles" at high frequencies. The 
plots in that figure are for a $10^7$ M$_\odot$ BH, so in the case of NGC4051, we would expect some wave-like 
pattern in the PSD above 10$^-3$ Hz or so. The residuals plot in Fig.\,\ref{figA1} do not show any significant 
wave-like residuals, however they could be of rather low amplitude (specially in the case of a spin zero BH, 
and a corona height larger than 10 R$_g$), so it may not be easy to detect them. As study of the expected 
PSD at low energies in the case of X-ray illumination of the disc is necessary to address quantitatively the 
X-ray reflection scenario.

\subsection{Implications for the nature of the variability of the X-ray corona}
A key finding of our study is that the bending frequency does not depend on energy in the 0.3-3.0 keV 
range (this study) and even up to 10 keV \citep{vaughan2011}. If the observed variability were due to 
changes in the hot corona, this would suggest that the hot corona in NGC~4051 maintains a constant 
temperature rather than increasing inward.
In the latter scenario, where accretion rate 
fluctuations propagate inwards \citep[e.g.][]{lyubarskii1997, kotov2001, arevalo2006}, we would expect the bending 
frequency to increase with increasing energy. For example, according to \cite{arevalo2006}, if the emissivity 
profile is such that \( E(r) \propto r^{-3} (1-\sqrt{r_{min}/r}) \), where \( r = R/R_s \), and if the bending 
frequency (in the PSD at a certain energy) corresponds to the viscous frequency at the radius where the photons 
in that band are mainly produced, then we would 
expect \( \frac{\nu_{bend} (3 \text{ keV})}{\nu_{bend}(0.3 \text{ keV})} \approx 3.5 \).If \( \nu_{\text{bend}} \approx 3 \times 10^{-4} \) 
at 0.3-0.5 keV (see Fig. 4), then we would expect \( \nu_{\text{bend}} \approx 1 \times 10^{-3} \) at 3 keV, 
which is highly improbable according to our results (see Fig. 4). 

Our work shows that the predictions of the propagation fluctuation model in its most basic form 
are not consistent with the observations. Propagation effects are likely to be more subtle, and 
an accreting multi-temperature corona may not be ruled out.  In any case, the 0.3-3keV band analysed 
here represents only a part of the broadband X-ray emission from the hot corona. Further studies 
of the PSDs at higher energy bands and in many more objects will be needed to reach a more detailed 
understanding of the energy dependence of the PSD and its comparison with the propagating fluctuations 
models.  

In the case of a single temperature corona, which is located within the inner disk, the bending frequency could 
correspond to the timescale of propagation of the sound wave in the radial direction, $t_{sound-r}$, at $R=R_{hot-out}$, 
where $R_{hot-out}$ is the outer radius of the hot corona and the inner radius of the accretion disk. At this radius 
sound waves may be generated and propagated inwards (within the hot corona) \citep[e.g.][]{cabanac2010} 
(sound waves will probably propagate within the disk as well, but perhaps damping may be very strong in 
this case). 
This  could explain that the characteristic bending timescale in the X-ray PSDs of NGC 4051 does not depend on energy. 
The frequency of the PSD break can be defined as  $\sim$ $c_s/R_{corona}$, 
where $c_s$ is the sound speed. For $c_s$ $\sim$ $3.1 \times 10^4 (T_{corona}/100~keV)^{1/2}$~Km/s, and $T_{corona}$ = 100~keV, 
and PSD break at $\sim$3.3$e^{-4}$~Hz,  we get $R_{corona}$ $\sim$30~R$_s$. This size is rather 
large and does not exactly agree with the size inferred by mirolensing and other 
studies \citep{reis2013, ursini2020, chartas2016}. We note that if the corona is highly 
magnetized, then other time scales may also be important and could be responsible for the 
observed bending frequencies like the Alfven and fast and slow magnetosonic wave speeds.  \\

NICER's exceptional time resolution has enabled the exploration of a new domain in the parameter 
space of X-ray variability in AGN, leading to the discovery of novel and unexpected phenomena. 
The observations allowed us to study the PSD in NGC~4051 at frequencies up to 0.1 Hz and and confirmed 
that the AGN corona has a uniform temperature. 
The study enhances our understanding of AGN coronae by providing evidence that the hot corona, rather 
than a warm one, is likely responsible for X-ray variability, suggesting a more stable and 
uniform structure. If this holds true for NGC 4051, it implies a relatively large corona size, 
challenging existing models and highlighting the need for more advanced theoretical models.
Gaining insight into AGN coronae is crucial not only for 
comprehending the central engines of AGN but also because these coronae could potentially be 
sites for the production of high-energy neutrinos \citep{icecube2022, padovani2024}.



\appendix

\begin{table*}
    \caption{XMM-Newton observation log for NGC 4051}
    \begin{tabular}{ccc}
    \hline
        Obs.\ ID & Obs.\ Date & Duration \\ 
        ~ & (YYYY-MM-DD) & (sec) \\ \hline
        0109141401 & 2001-05-16 & 121958 \\ 
        0157560101 & 2002-11-22 & 51866 \\ 
        0606320101 & 2009-05-03 & 45717 \\ 
        0606320201 & 2009-05-05 & 45645 \\ 
        0606320301 & 2009-05-09 & 45548 \\ 
        0606320401 & 2009-05-11 & 45447 \\ 
        0606321301 & 2009-05-15 & 32644 \\ 
        0606321401 & 2009-05-17 & 42433 \\ 
        0606321501 & 2009-05-19 & 41813 \\ 
        0606321601 & 2009-05-21 & 41936 \\ 
        0606321701 & 2009-05-27 & 44919 \\ 
        0606321801 & 2009-05-29 & 43726 \\ 
        0606321901 & 2009-06-02 & 44891 \\ 
        0606322001 & 2009-06-04 & 39756 \\ 
        0606322101 & 2009-06-08 & 43545 \\ 
        0606322201 & 2009-06-10 & 44453 \\ 
        0606322301 & 2009-06-16 & 42717 \\ 
        0830430201 & 2018-11-07 & 83200 \\ 
        0830430801 & 2018-11-09 & 85500 \\ \hline
    \end{tabular}\\
    The first column contains the Observation identifier. The second and last 
    columns display the start date and duration of the respective observations.
    \label{table_Xmm}
\end{table*}

\begin{figure*}
\center
\includegraphics[scale=0.34, angle =-90, trim=0 33 50 0, clip]{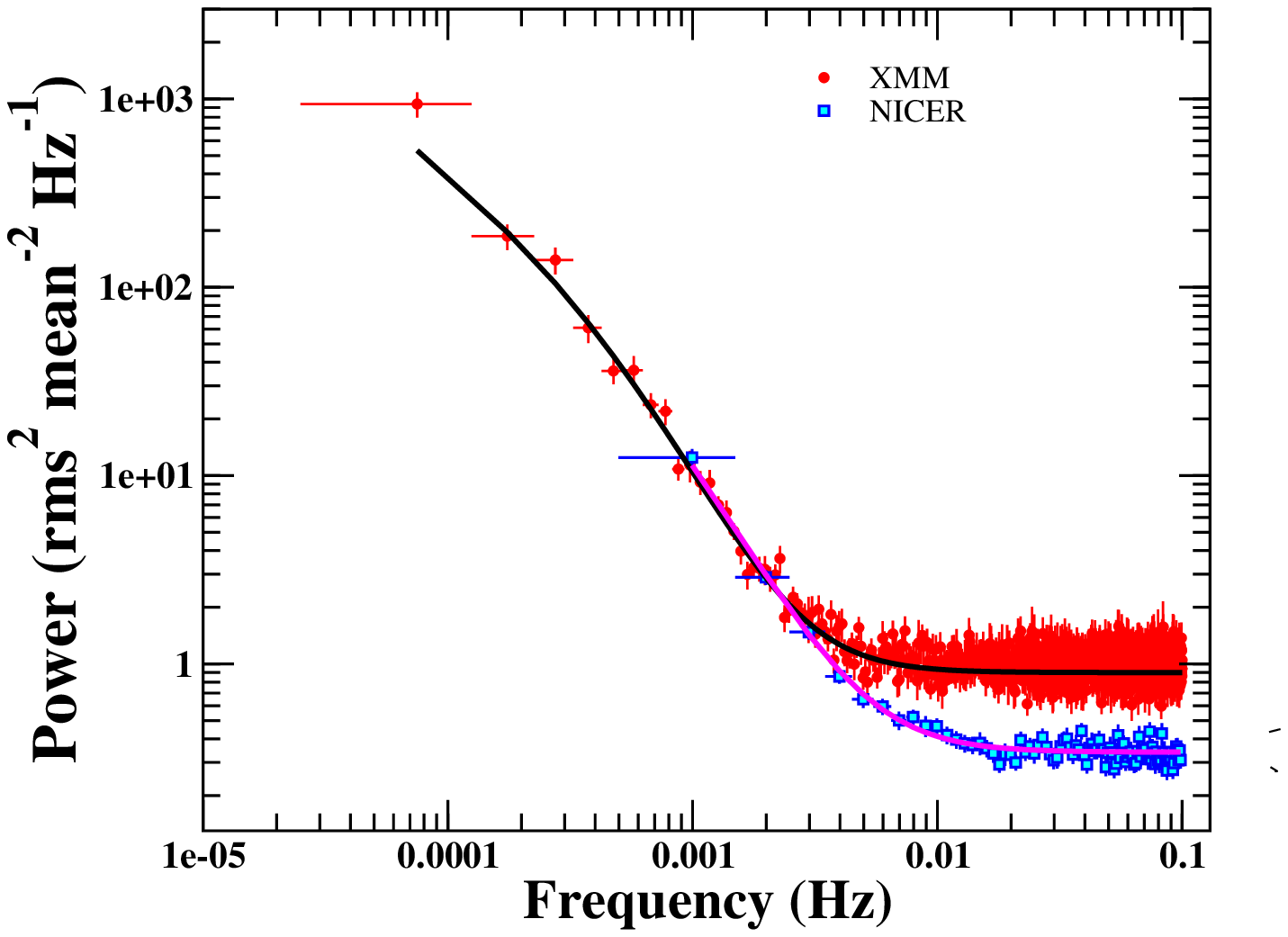}
\includegraphics[scale=0.31, angle=-90, trim=30 0.2 20 0, clip]{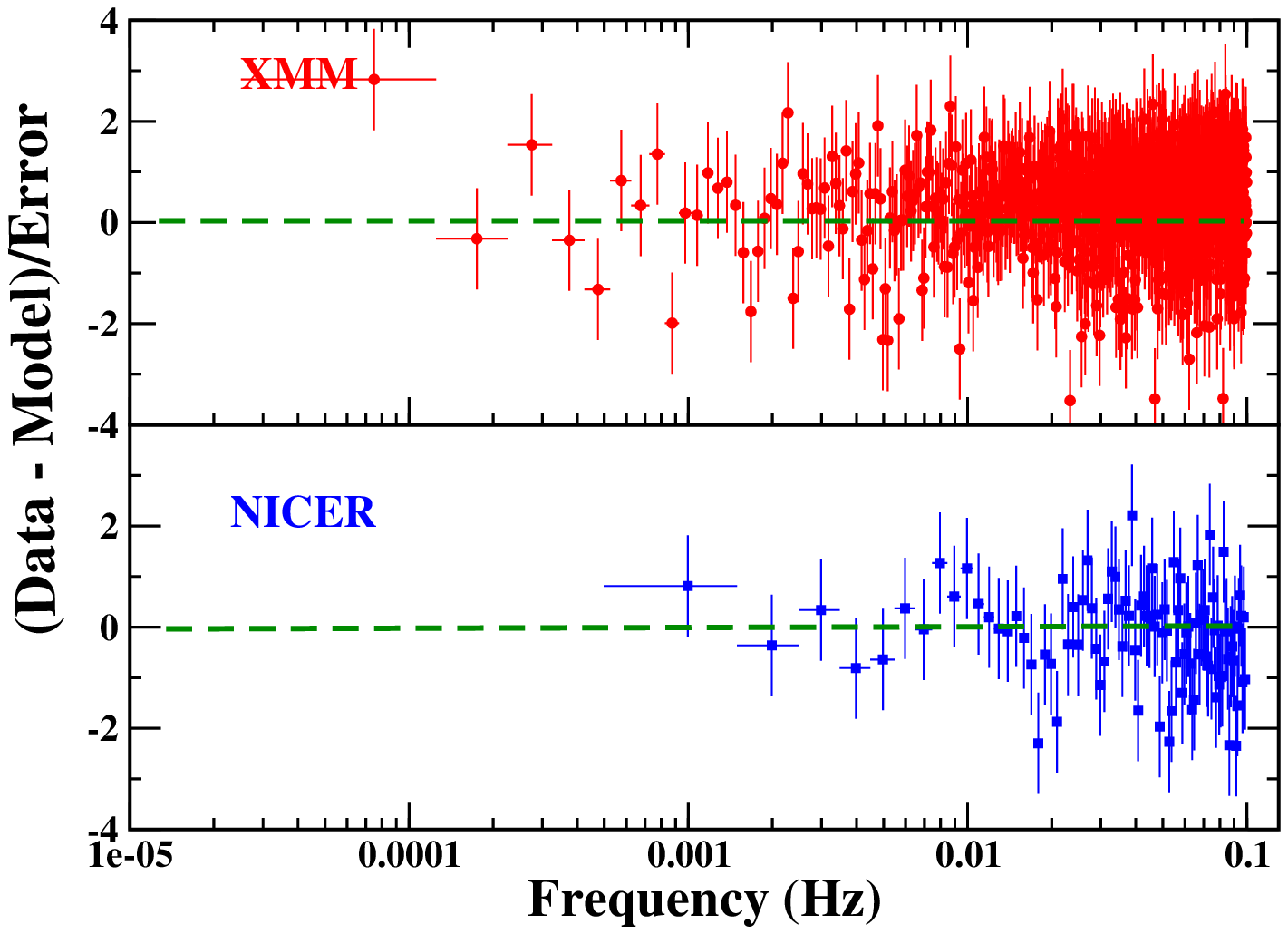}
\caption{PSDs in the 0.3-0.5 keV range (same as Figure 2). Best-fit models are plotted in the case when $\alpha_{high}$ is separate for the XMM and NICER PSDs.}
\label{figA1}
\end{figure*}

\begin{figure*}
\center
\includegraphics[scale=0.34, angle =-90, trim=0 33 50 0, clip]{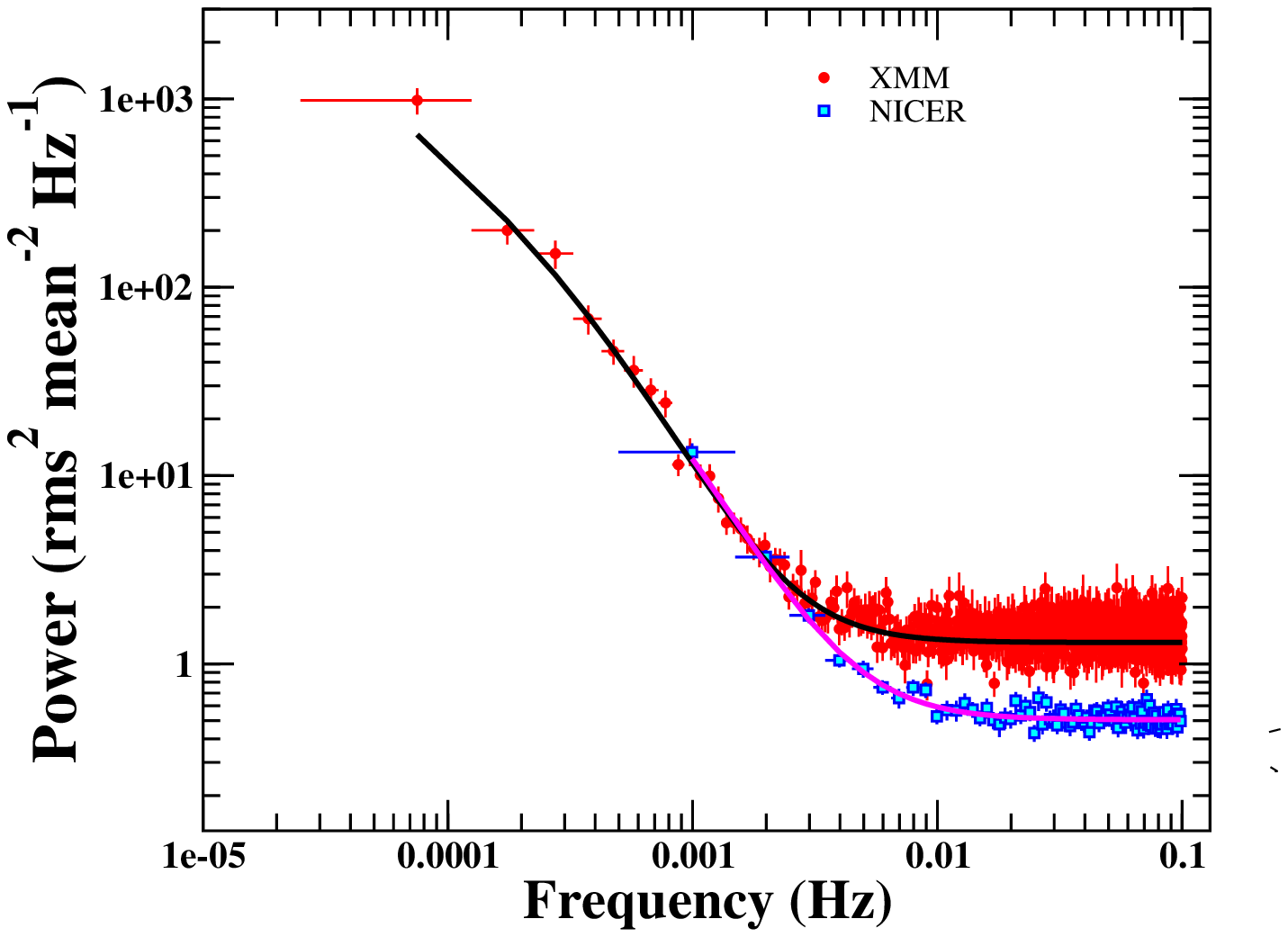}
\includegraphics[scale=0.31, angle=-90, trim=30 0.2 20 0, clip]{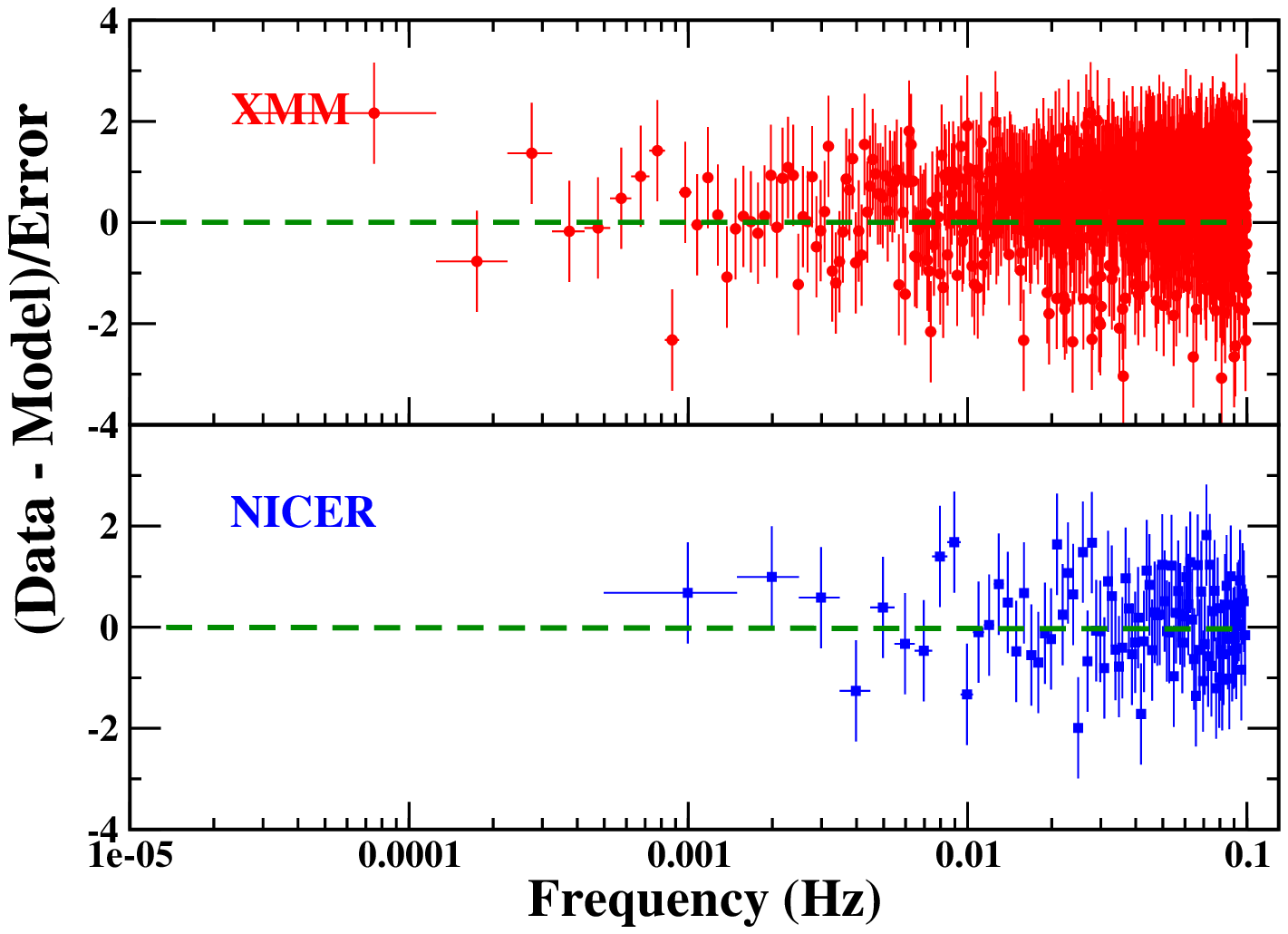}
\caption{PSDs in the 0.5-0.7 keV range (same as Fig.\,\ref{figA1}). }
\label{figA2}
\end{figure*}

\begin{figure*}
\center
\includegraphics[scale=0.34, angle =-90, trim=0 33 50 0, clip]{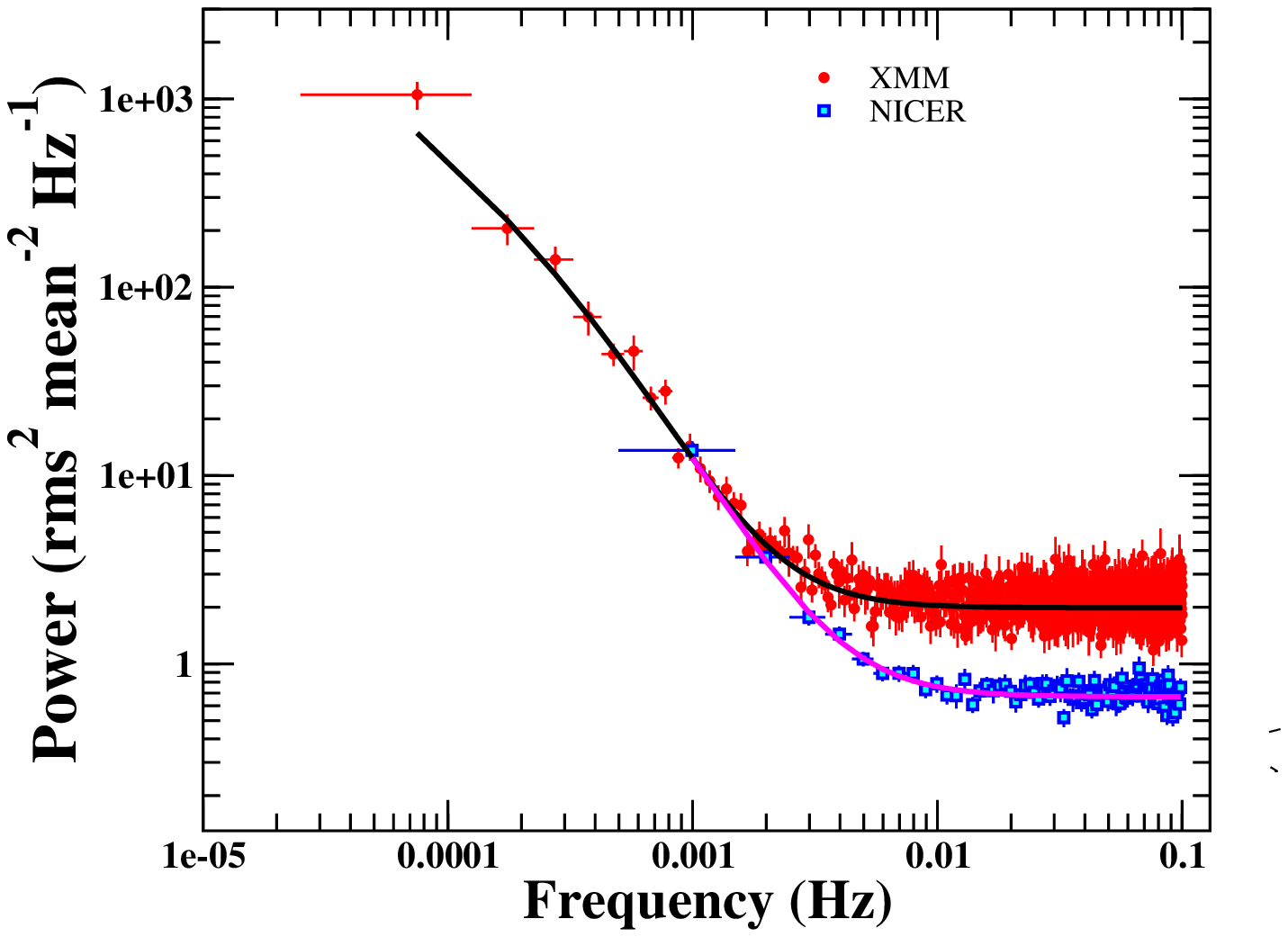}
\includegraphics[scale=0.31, angle=-90, trim=30 0.2 20 0, clip]{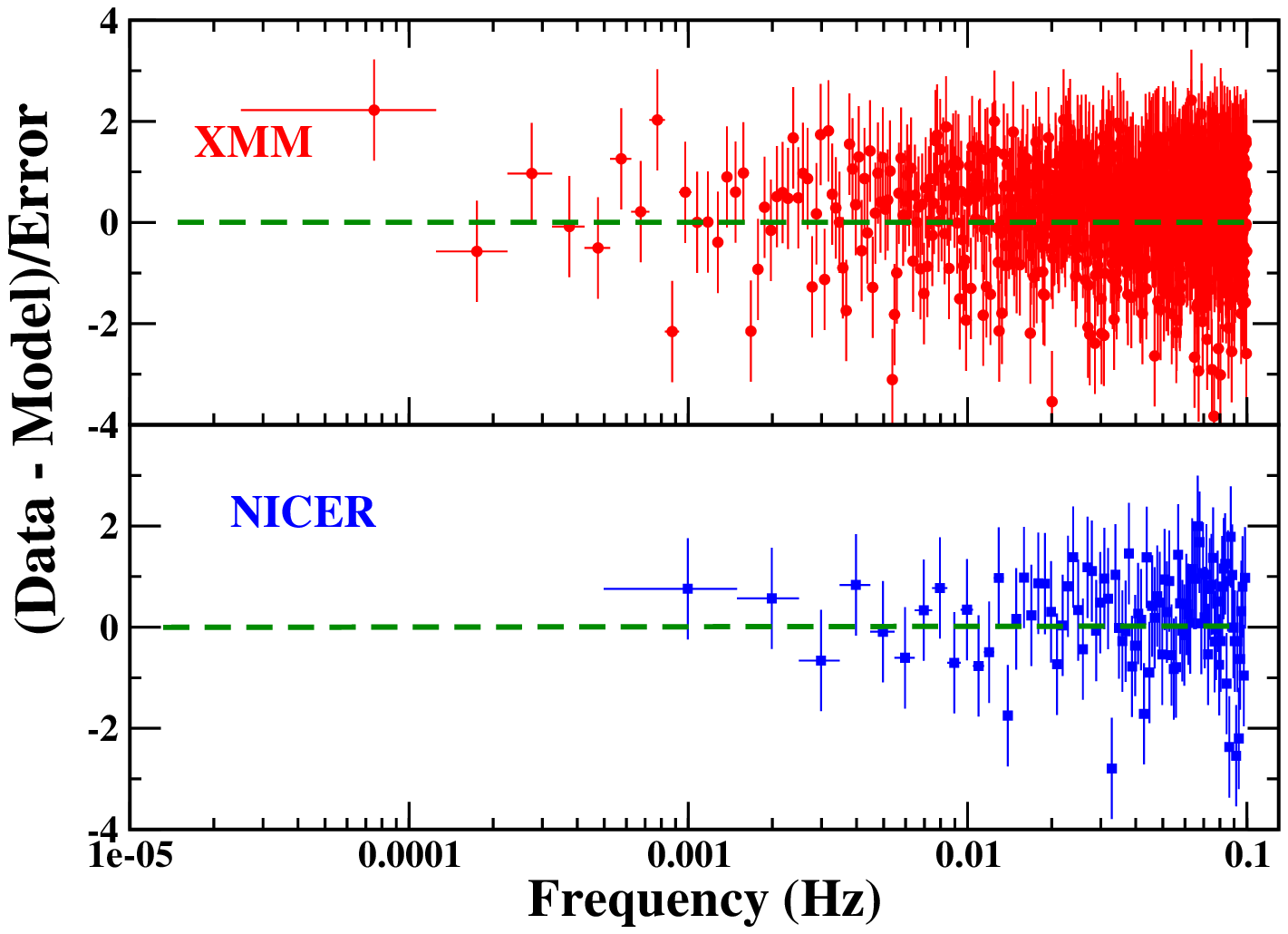}
\caption{PSDs in the 0.7-1.0 keV range (same as Fig.\,\ref{figA1}). }
\label{figA3}
\end{figure*}

\begin{figure*}
\center
\includegraphics[scale=0.34, angle =-90, trim=0 33 50 0, clip]{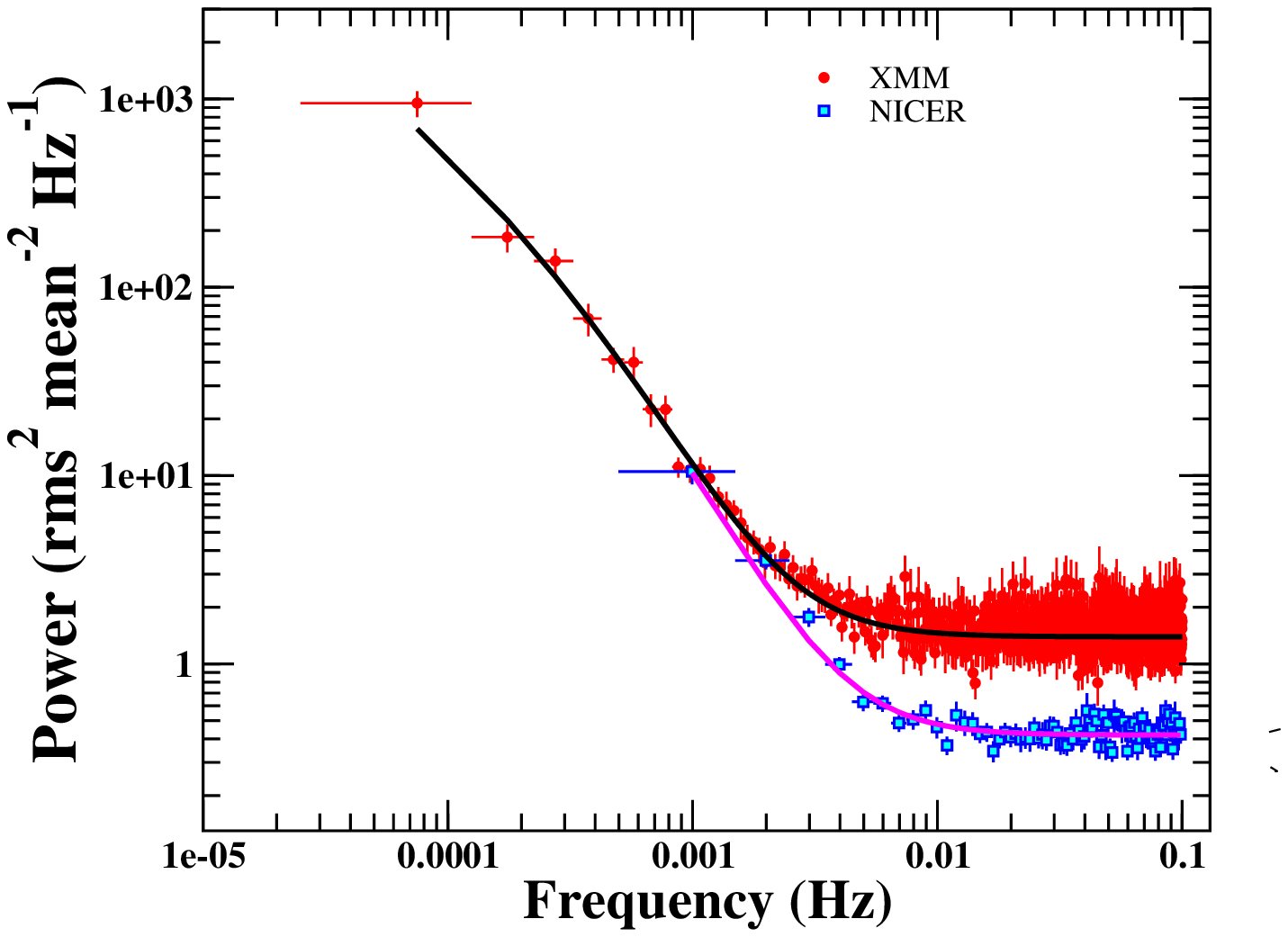}
\includegraphics[scale=0.31, angle=-90, trim=30 0.2 20 0, clip]{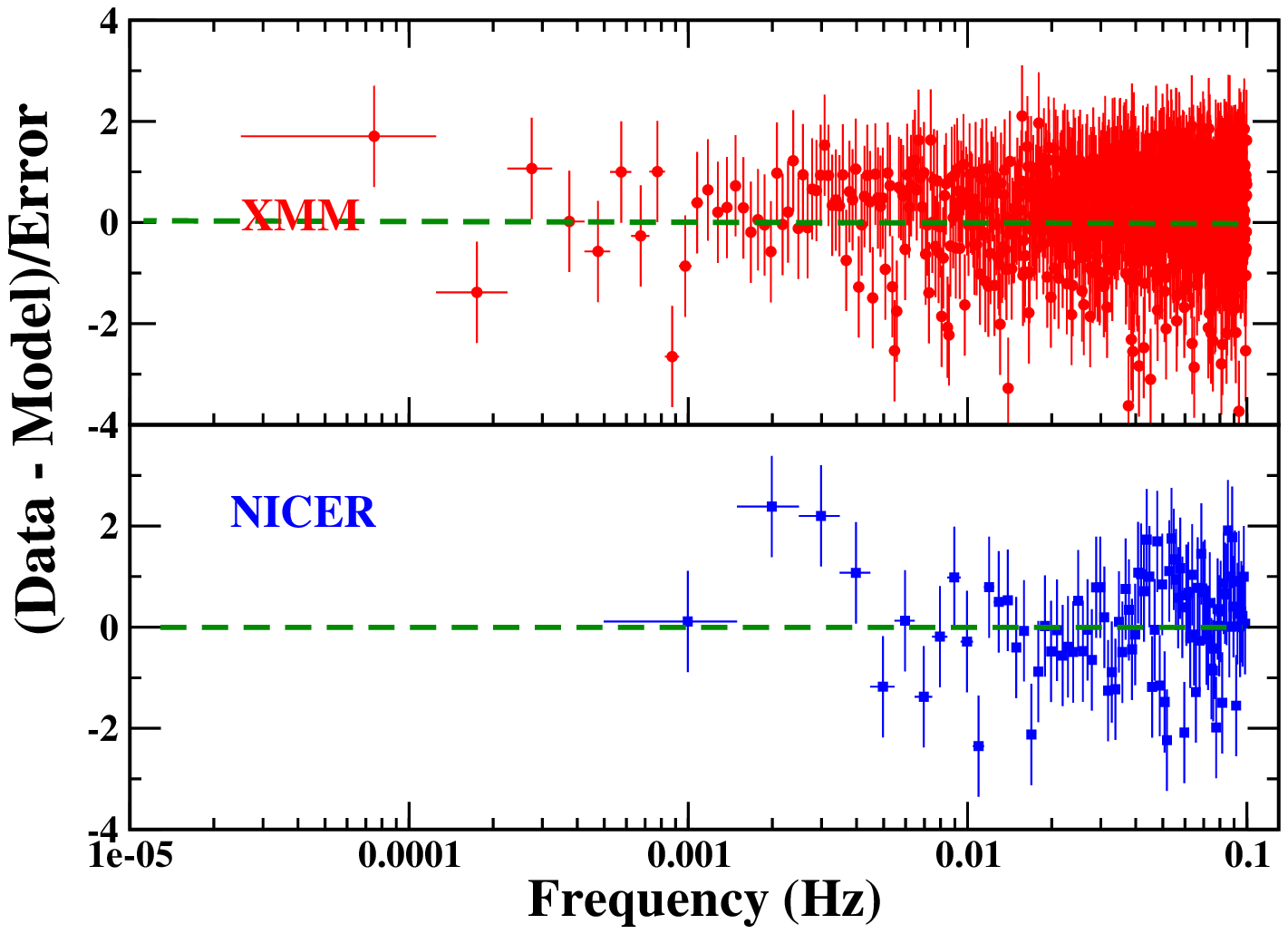}
\caption{PSDs in the 1.0-3.0 keV range (same as Fig.\,\ref{figA1}). }
\label{figA4}
\end{figure*}

\end{document}